\documentclass[11pt]{article}
\usepackage{graphicx}
\usepackage{amsmath}
\usepackage{amssymb}
\usepackage{amsxtra}

\title{The cosmic $e^\pm$ anomaly}
\author{Katie Auchettl and Csaba Bal\'azs \\
Monash Centre for Astrophysics and \\
ARC Centre of Excellence for Particle Physics at the Tera-scale \\
School of Physics, Monash University, Victoria 3800 Australia}

\begin{document}
\maketitle

\begin{abstract}
Via a Bayesian likelihood analysis using 219 recent cosmic ray spectral data points we extract the anomalous part of the cosmic $e^\pm$ flux.  First we show that a significant tension exists between the $e^\pm$ related and the rest of the fluxes.  Interpreting this tension as the presence of an anomalous component in the $e^\pm$ related data, we then infer the values of selected cosmic ray propagation parameters excluding the anomalous data sample from the analysis.  Based on these values we calculate background predictions with theoretical uncertainties for PAMELA and Fermi-LAT.  We find a statistically significant deviation between the Fermi-LAT $e^-+e^+$ data and the predicted background even when (systematic) uncertainties are taken into account.  Identifying this deviation as an anomalous $e^\pm$ contribution, we make an attempt to distinguish between various sources that may be responsible for the anomalous $e^\pm$ flux. 
\end{abstract}

\section{Introduction}\label{sec.intro} 

 
Cosmic ray observations provided significant and puzzling deviations from theoretical predictions over the last decades.  Experiments such as 
TS \cite{Golden:1992zm}, 
AMS \cite{Alcaraz:2000bf}, 
CAPRICE \cite{Boezio:2001ac}, 
MASS \cite{Grimani:2002yz}, and
HEAT \cite{Barwick:1997ig, Beatty:2004cy}
established a hint of an excess of high energy electrons and/or positrons.  
Measurements of the PAMELA satellite confirmed these results by finding an excess over the theoretical predictions in the $e^-/(e^- + e^+)$ flux for $E > 10$ GeV \cite{Adriani:2008zr}.  The PAMELA data seem to deviate from the theoretical predictions even when experimental and theoretical uncertainties are taken into account \cite{Timur:2011vv}. 
An excess in the $e^- + e^+$ flux was also found by 
AMS \cite{Aguilar:2002ad}, 
PPB-BETS \cite{Torii:2008xu}, and 
HESS \cite{Aharonian:2008aa,Aharonian:2009ah}. 
Recently the Fermi-LAT satellite confirmed the $e^- + e^+$ excess above 100 GeV \cite{Ackermann:2010ij}.  The deviation between the Fermi-LAT data and the theoretical $e^- + e^+$ prediction is significant.  This deviation was recently confirmed by the PAMELA collaboration which found the $e^-$ flux to be consistent with the Fermi-LAT data.

Many attempts were to explain the deviation between the data and theory.  New physics was invoked ranging from modification of the cosmic ray propagation, through supernova remnants, to dark matter annihilation.  Ref. \cite{Serpico:2011wg} summarizes these speculations.  
%
%
Whether the $e^\pm$ anomaly exists depends on the theoretical prediction of the cosmic ray background.  The theoretical prediction is challenging because of the lack of precise knowledge of the cosmic ray sources, and because the cosmic ray propagation model has numerous free parameters, such as convection velocities, spatial diffusion coefficients and momentum loss rates. 


Motivated by traces of possible new physics in the Fermi-LAT data, we attempt to determine the size of the anomalous contribution in the cosmic $e^\pm$ flux.  Our method involves the following steps.
First we find the cosmic ray propagation parameters that influence the $e^\pm$ flux measured by Fermi-LAT and PAMELA the most.  
Then we subject the cosmic ray data, other than the Fermi-LAT and PAMELA $e^\pm$ measurements to a Bayesian likelihood analysis, to determine the preferred values and the 68 \% (1-$\sigma$) credibility regions of the relevant propagation parameters.
Based on the central values and 1-$\sigma$ credibility regions of these propagation parameters we then predict the background flux, with uncertainties, for Fermi-LAT and PAMELA.
Finally, we extract the anomalous part of the spectrum by subtracting the background prediction from the Fermi-LAT and PAMELA measurement.

\section{Theory of cosmic ray propagation}

Cosmic ray propagation is described by the diffusion model \cite{Ginzburg:1990sk}.  This model assumes homogeneous propagation of charged particles within the Galactic disk and it also takes into account cooling effects.  The phase-space density $\psi_a (\vec r,p,t)$ of a particular cosmic ray species at a Galactic radius of $\vec r$ can be calculated solving the transport equation which has the general form \cite{Strong:2007nh}
\begin{eqnarray}
 {\partial \psi_a (\vec r,p,t) \over \partial t} 
 & = &
 Q_a(\vec r, p, t) 
 + \nabla \cdot ( D_{xx}\nabla\psi_a - \vec V\psi_a ) 
 - ({1\over\tau_f} + {1\over\tau_r})\psi_a \nonumber \\
 & + & {\partial\over\partial p}\, \left( p^2 D_{pp} {\partial\over\partial p}\, {1\over p^2}\, \psi_a \right)  
 - {\partial\over\partial p} \left( \dot{p} \psi_a
 - {p\over 3} \, (\nabla \cdot \vec V )\psi_a \right) .
\label{eq:transport}
\end{eqnarray}

Here $q(\vec r, p, t)$ is the source term of primary and secondary cosmic ray contributions.  The spatial diffusion coefficient $D_{xx}$ has the form
\begin{eqnarray}
\label{eq:Dxx}
D_{xx} =  D_{0xx} \beta \left({R\over {\rm GeV}}\right)^\delta ,
\end{eqnarray}
where $\beta = v/c$, and $R=pc/eZ$ is the magnetic rigidity of the particles which describes a particle's resistance to deflection by a magnetic field.  Here $Z$ is the effective nuclear charge of the particle, $v$ is its velocity, $p$ is its momentum, $e$ is its charge, and $c$ is the speed of light.  
The exponent $\delta$ indicates the power law dependence of the spatial diffusion coefficient $D_{xx}$.  

Diffusion in momentum space (diffusive re-acceleration) is described by the coefficient $D_{pp}$
\begin{eqnarray}
\label{eq:DppDxx}
D_{pp} D_{xx} = {4 p^2 {v_A}^2\over 3\delta(4-\delta^2)(4-\delta) w}\ .
\end{eqnarray}
Here $v_A$ is the Alfven speed, the parameter $w$ characterises the level of hydromagnetic turbulence experienced by the cosmic rays in the interstellar medium \cite{1994ApJ...431..705S}.  
In Eq.(\ref{eq:transport}), $\vec V$ is the convection velocity and the parameter $\tau_f$ is the time-scale of the fragmentation loss, and $\tau_r$ is the radioactive decay time-scale. 

The GalProp numerical package solves the propagation equation numerically for $Z \geq 1$ nuclei, as well as for electrons and positrons on a two dimensional spatial grid with cylindrical symmetry in the Galaxy \cite{Strong:2007nh}.  The input parameter file for GalProp has a number of free parameters which are available for the author to define.  These can be classified into a number of subsets: the diffusion of cosmic ray, the primary cosmic ray sources and radiative energy losses of these primary cosmic rays.  The diffusion subset is described by the parameters defined above: 
\begin{eqnarray}
 D_{0xx}, \delta, L, v_A, \partial {\vec V} / \partial z . 
\end{eqnarray}
The most relevant parameters in the primary cosmic ray source subset are: 
\begin{eqnarray}
 R_{ref}^{e^-}, \gamma_1^{e^-}, \gamma_2^{e^-}, R_{ref}^{nucleus}, \gamma_1^{nucleus}, \gamma_2^{nucleus} .
\end{eqnarray}
Here $\gamma_1^{e^-}$ and $\gamma_2^{e^-}$ are primary source electron injection indices.  They specifying the steepness of the electron injection spectrum, $dq(p)/dp \propto p^{\gamma_i^{e^-}}$, below and above of a reference rigidity $R_{ref}^{e^-}$.  There are separate injection indices for nuclei defined by $\gamma_1^{nucleus}$ and $\gamma_2^{nucleus}$ below and above $R_{ref}^{nucleus}$.  For further details we refer the reader to \cite{Strong:2007nh}.

\section{Parameter space and uncertainties}

To reduce the dimension of the parameter space we tested the robustness of the $e^\pm$ flux against the variation of nearly all individual parameters and found that it is mostly sensitive to the following propagation parameters:
\begin{eqnarray} 
\label{eq:P}
P = \{\gamma_1^{e^-}, \gamma_2^{nucleus}, \delta_1, \delta_2, D_{0xx} \} .
\end{eqnarray}
Here $\gamma_1^{e^-}$ and $\gamma_2^{nucleus}$ are the primary electron and nucleus injection indices, $\delta_1$ and $\delta_2$ are spatial diffusion coefficients below and above a reference rigidity $\rho_0$, and $D_{0xx}$ determines the normalization of the spatial diffusion coefficient.

Our calculations confirmed the findings of a recent study by \cite{Cotta:2010ej} that the $e^\pm$ flux is sensitive to the change of the Galactic plane height $L$.  Indeed \cite{1994ApJ...431..705S} have shown that there is a connection between $L$ and $D_{0xx}$: 
\begin{eqnarray} 
\label{eq:D0xx}
D_{0xx} = \frac{2 c (1-\delta) L^{1-\delta}}{3 \pi w \delta (\delta + 2)} .
\end{eqnarray}
Thus, varying the cylinder height amounts to the redefinition of $D_{0xx}$ as also noticed by Ref. \cite{DiBernardo:2010is}.  In the light of this, we fix $L$ to 4 kpc and use $D_{0xx}$ as free parameter. 


We treat the normalizations of the $e^-$, $e^+$, ${\bar {\rm p}}$/p, B/C, (SC+Ti+V)/Fe and Be-10/Be-9 fluxes as theoretical nuisances parameters.  
\begin{eqnarray} 
\label{eq:Pn}
P_{nuisance} = \{ \Phi_{e^-}^0, \Phi_{e^+}^0, \Phi_{{\bar p}/p}^0, \Phi_{B/C}^0, \Phi_{(SC+Ti+V)/Fe}^0, \Phi_{Be-10/Be-9}^0 \} .
\end{eqnarray}


When evaluating the uncertainties, following \cite{Trotta:2010mx}, we ignore theoretical uncertainties and combine statistical and systematic experimental uncertainties in quadrature
\begin{eqnarray}
 \label{eq:sigma1}
 \sigma_i^2 = \sigma_{i, statistical}^2 + \sigma_{i, systematic}^2.
\end{eqnarray}
This can be done for Fermi-LAT and the latest PAMELA $e^-$ flux.  Unfortunately, systematic uncertainties are not available for the rest of the cosmic ray measurements.  When this is the case, as an estimate of the systematics, we define $\sigma_i$ as the rescaled statistical uncertainty
\begin{eqnarray}
 \label{eq:sigma2}
 \sigma_i^2 = \sigma_{i, statistical}^2/\tau_i .
\end{eqnarray}
For simplicity, in this study, we use the same scale factor $\tau_i$ for all data points where systematic uncertainty is not available.  To remain mostly consistent with the work of \cite{Trotta:2010mx}, we set this common scale factor to a conservative value that they use: $\tau_i = 0.2$.  We checked that our conclusions only mildly depend on this choice.  Further details about our Bayesian parameter inference can be found in \cite{Auchettl:2011wi}.

\section{Experimental data}
\label{sec:ExpDat}

We included 219 of the most recent experimental data points in our statistical analysis.  These contained 114 $e^\pm$ related, and 105 $\bar{p}/p$, B/C, (Sc+Ti+V)/Fe and Be-10/Be-9 cosmic ray flux measurements.  These data are summarized in Table \ref{tab:data}.

\begin{table}[h]
\begin{center}
\caption{Cosmic ray experiments and their energy ranges over which we have chosen the data points for our analysis.  We split the data into two groups: $e^\pm$ flux related (first five lines in the table), and the rest.  We perform two independent Bayesian analyses to show the significant tension between the two data sets.
\vspace{3mm}
\label{tab:data}}
\begin{tabular}{lllc}
\hline
\hline
\bf Measured flux  & \bf Experiment                                & \bf Energy   & \bf Data   \\
                   &                                               & (GeV)        & \bf points \\
\hline
\hline
                   & AMS \cite{Aguilar:2002ad}                     & 0.60 - 0.91  & 3  \\
 $e^+ + e^-$       & Fermi-LAT \cite{Ackermann:2010ij}             & 7.05 - 886   & 47 \\
                   & HESS \cite{Aharonian:2008aa,Aharonian:2009ah} & 918 - 3480   & 9  \\
\hline                                                                             
 $e^+/(e^+ + e^-)$ & PAMELA \cite{Adriani:2010ib}                  & 1.65 - 82.40 & 16 \\
\hline    
 $e^-$             & PAMELA \cite{Adriani:2011xv}                  & 1.11 - 491.4 & 39 \\
\hline
\hline                                                                             
 $\bar{p}$/p       & PAMELA \cite{Adriani:2010rc}                  & 0.28 - 129   & 23 \\
\hline                                                                            
                   & IMP8 \cite{Moskalenko:2001ya}                 & 0.03 - 0.11  &  7 \\
                   & ISEE3 \cite{1988ApJ...328..940K}              & 0.12 - 0.18  &  6 \\
 B/C               & Lezniak et al. \cite{1978ApJ...223..676L}     & 0.30 - 0.50  &  2 \\
                   & HEAO3 \cite{1990AA...233...96E}               & 0.62 - 0.99  &  3 \\
                   & PAMELA \cite{PAMELABORONCARBON}               & 1.24 - 72.36 &  8 \\
                   & CREAM \cite{Ahn:2008my}                       &   91 - 1433  &  3 \\
\hline                                                                             
 (Sc+Ti+V)/Fe      & ACE \cite{2000AIPC..528..421D}                & 0.14 - 35    & 20 \\
                   & SANRIKU \cite{1999ICRC....3..105H}            & 46 - 460     &  6 \\
\hline                                                                             
                   & Wiedenbeck et al. \cite{1980ApJ...239L.139W}   & 0.003 - 0.029 &  3 \\
                   & Garcia-Munoz et al. \cite{1981ICRC....9..195G} & 0.034 - 0.034 &  1 \\
                   & Wiedenbeck et al. \cite{1980ApJ...239L.139W}   & 0.06 - 0.06   &  1 \\
 Be-10/Be-9        & ISOMAX98 \cite{2001ICRC....5.1655H}             & 0.08 - 0.08   &  1 \\
                   & ACE-CRIS \cite{davis:421}                       & 0.11 - 0.11   &  1 \\
                   & ACE \cite{2001AdSpR..27..727Y}                  & 0.13 - 0.13   &  1 \\
                   & AMS-02 \cite{2004EPJC...33S.941B2}              & 0.15 - 9.03   & 15 \\
\hline                                                                         
\hline
\end{tabular}
\end{center}
\end{table}

\section{Is there a cosmic ray anomaly?}


In this section we investigate whether the present cosmic ray data justify the existence of an anomaly in the $e^\pm$ spectrum.  To this end we divide the cosmic ray data into two groups: 114 measurements containing observations of $e^\pm$ fluxes (AMS, Fermi, HESS, and PAMELA) and the rest of 105 data points ($\bar{p}/p$, B/C, (Sc+Ti+V)/Fe, Be-10/Be-9).  We perform a Bayesian analysis independently on these two sets of data extracting the preferred values of the propagation parameters.  


\begin{figure}
\begin{center}
\includegraphics[width=0.45\textwidth]{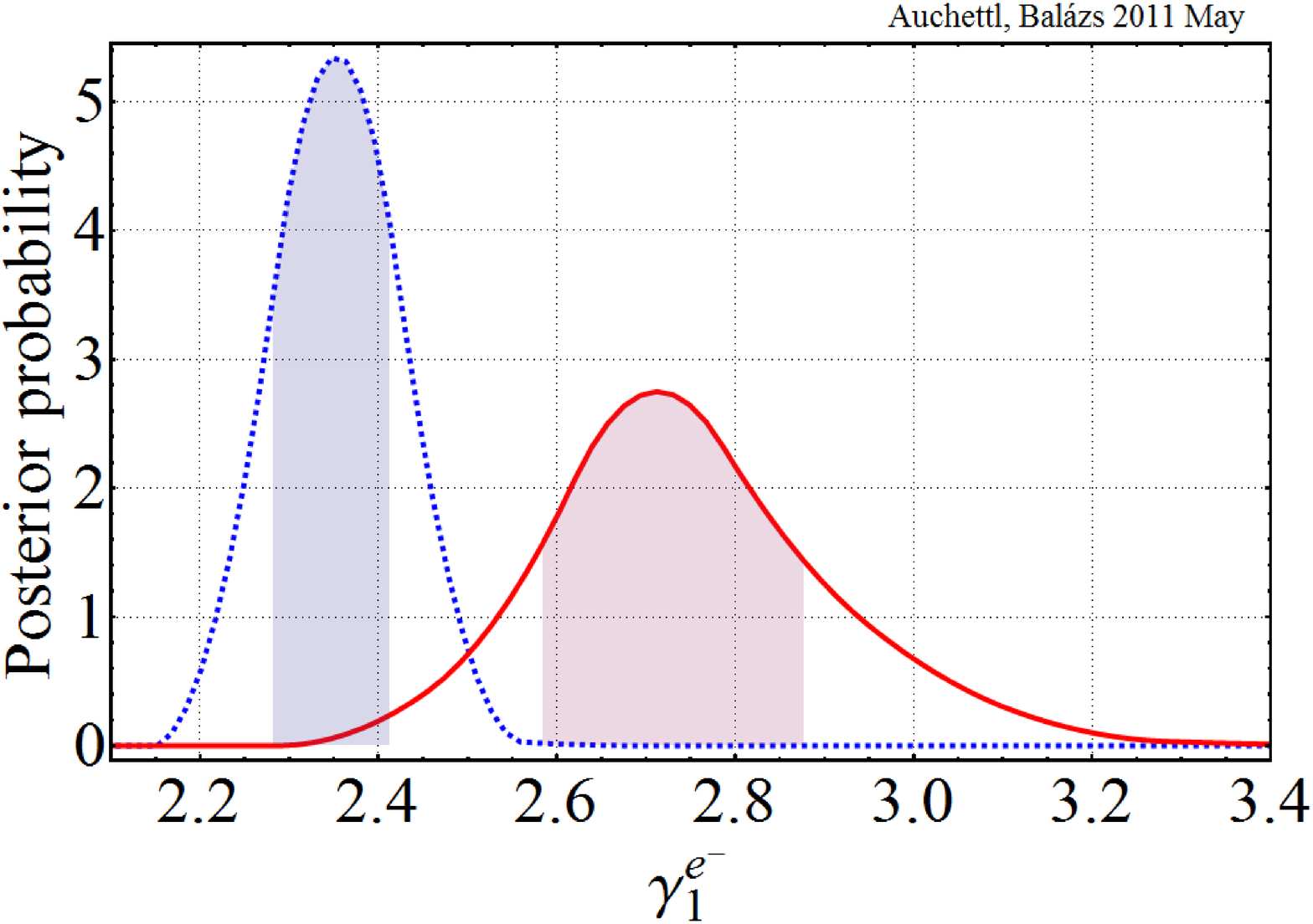}
\includegraphics[width=0.45\textwidth]{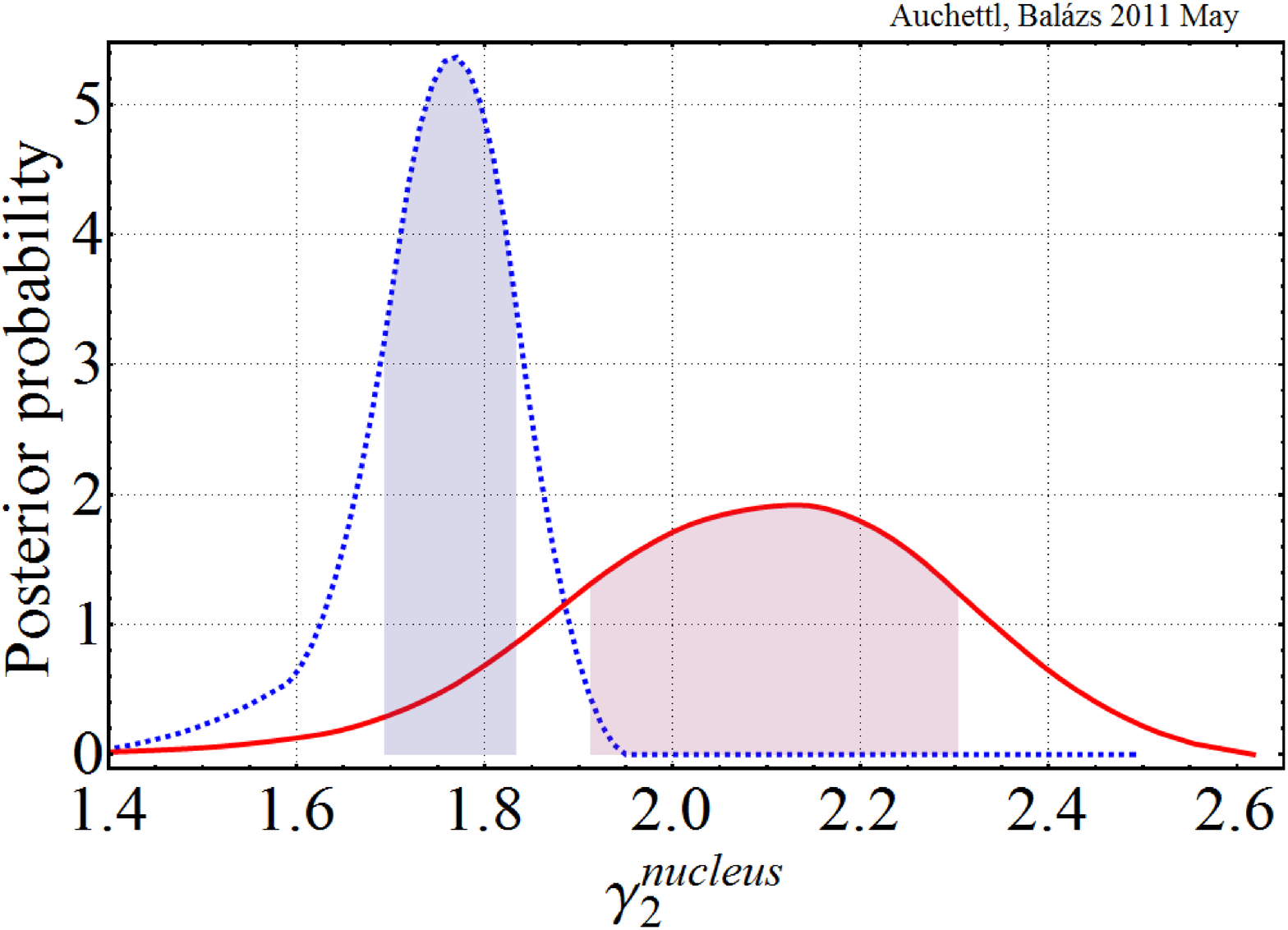}
\includegraphics[width=0.45\textwidth]{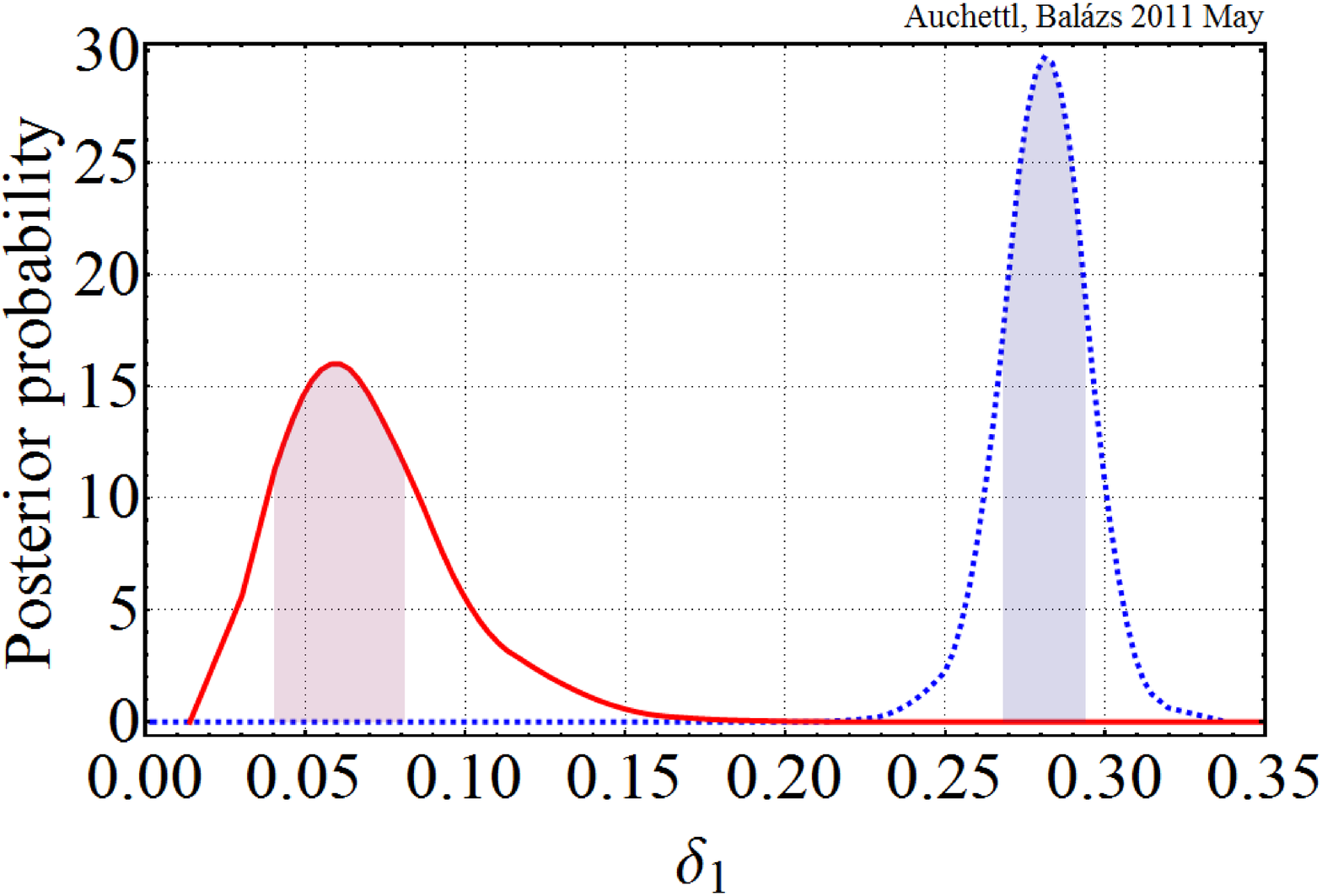}
\includegraphics[width=0.45\textwidth]{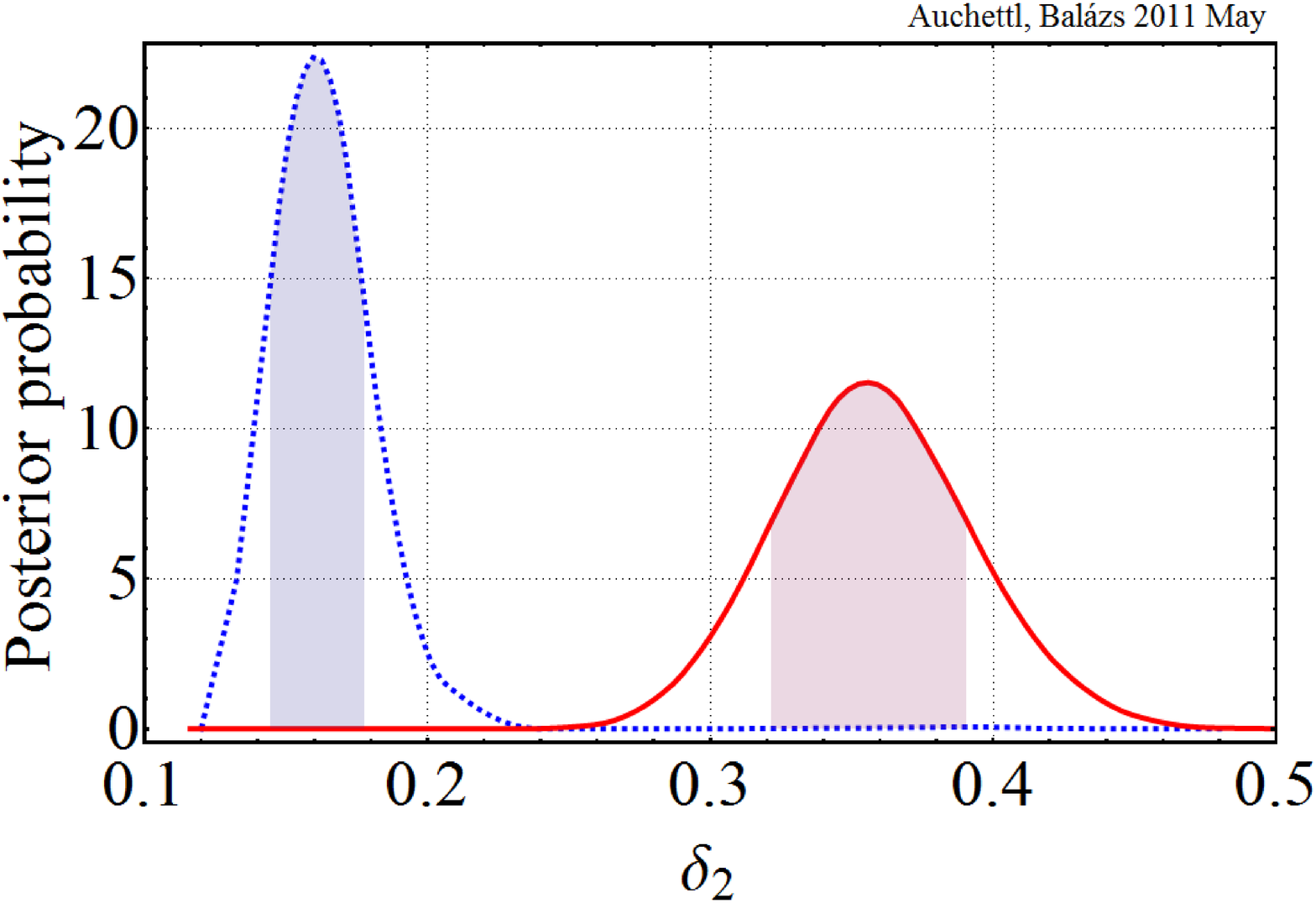}
\end{center}
\caption{Marginalized posterior probability distributions of propagation parameters listed in Eq.(\ref{eq:P}).  The dashed blue curves show results with likelihood functions containing $e^\pm$ flux data while the likelihood functions for the solid red curves contain only the rest of the comic ray data.  Shaded areas show the 68 \% credibility regions.  A statistically significant tension between the $e^\pm$ and the rest of the data is evident in the lower frames.}
\label{fig:Posteriors}
\end{figure}


Fig. \ref{fig:Posteriors} clearly shows that the two subsets of cosmic ray data are inconsistent with the hypothesis that the cosmic ray propagation model and/or sources implemented in GalProp provides a good theoretical description.  
Our interpretation of the tension between the $e^\pm$ fluxes and the rest of the cosmic ray data is that the measurements of PAMELA and Fermi-LAT are affected by new physics which is unaccounted for by the propagation model and/or cosmic ray sources included in our calculation.

\section{The size of the $e^\pm$ anomaly}

\begin{figure}
\begin{center}
\includegraphics[width=0.45\textwidth]{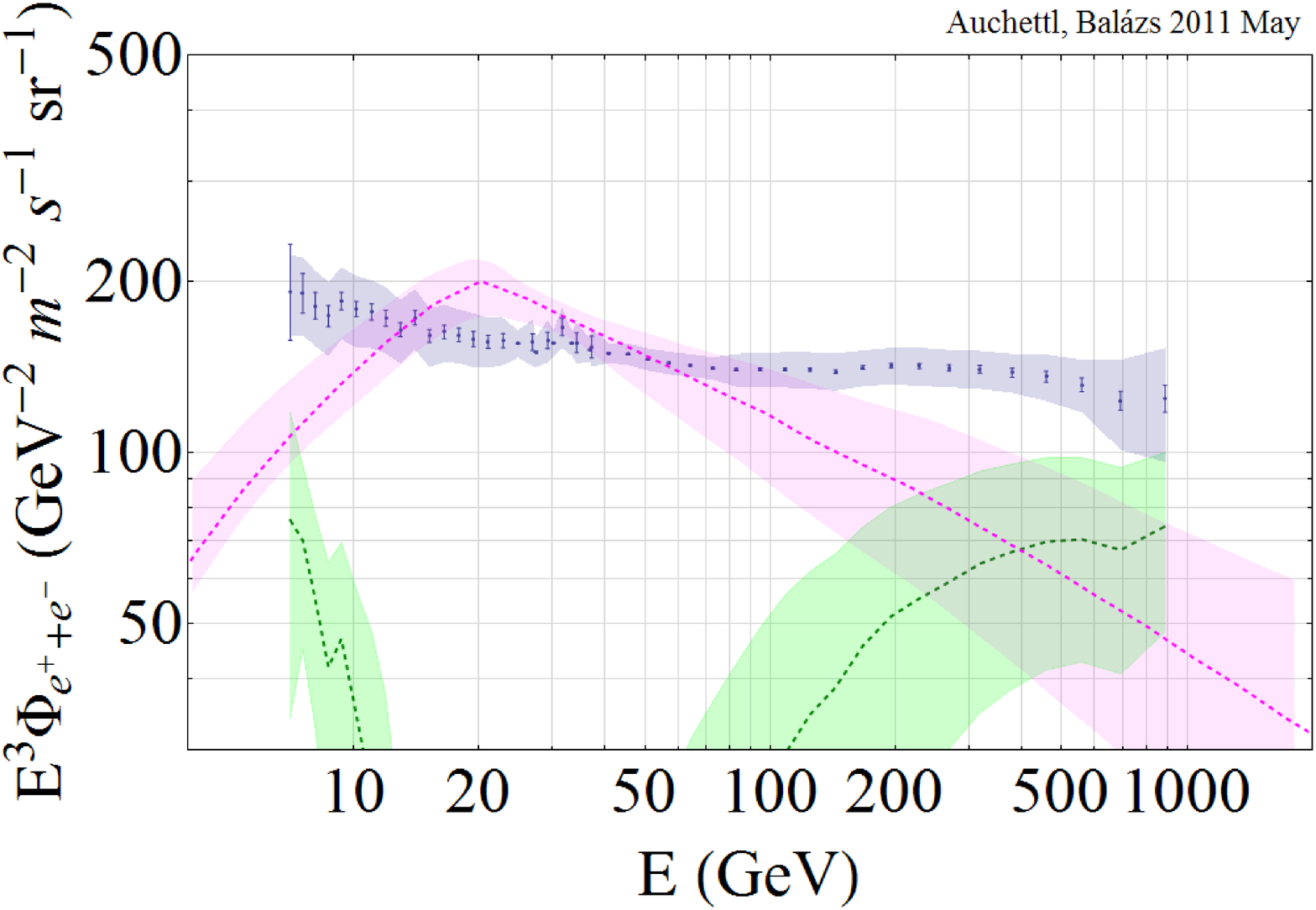}
\includegraphics[width=0.45\textwidth]{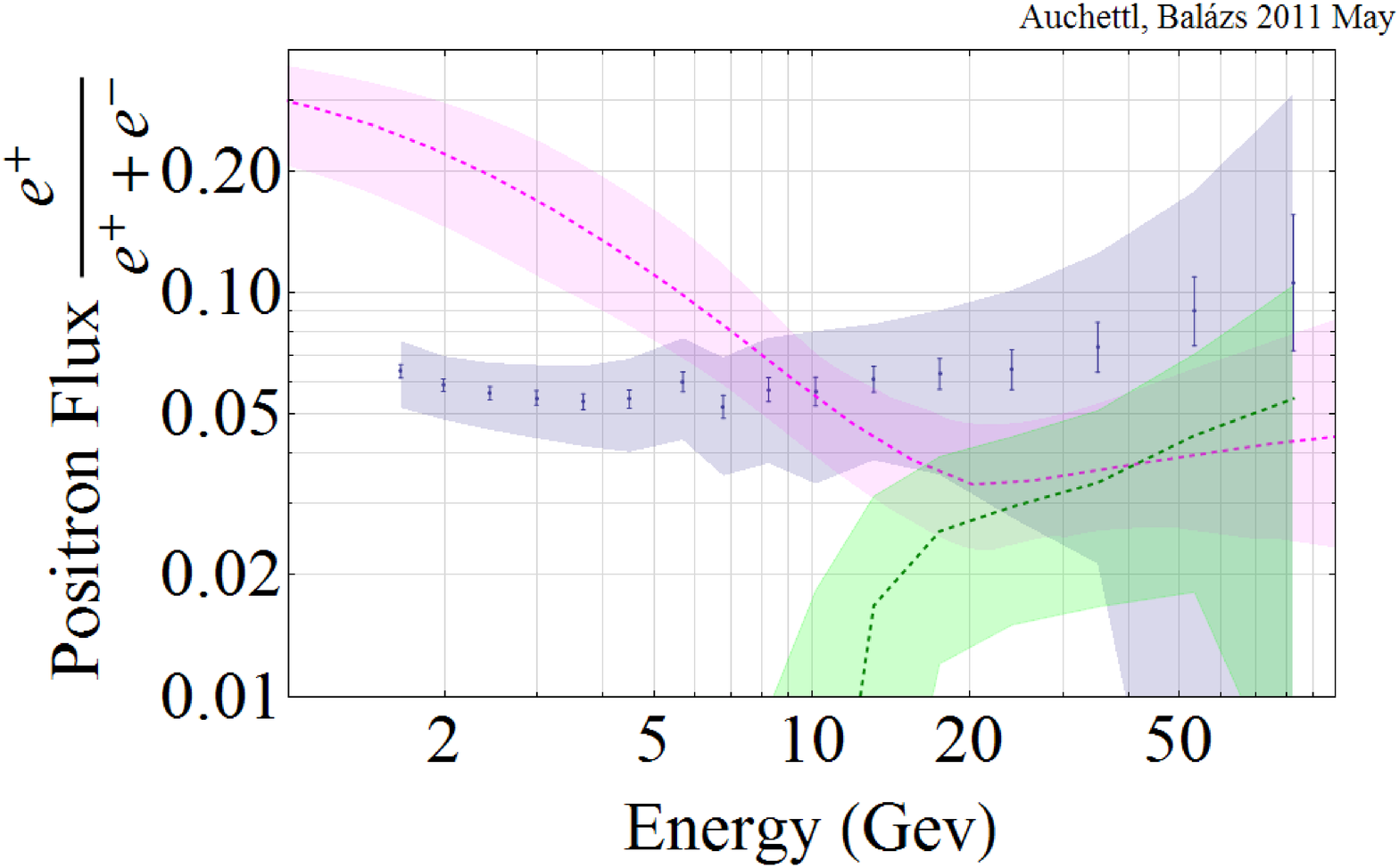}
\end{center}
\caption{Electron-positron fluxes measured by Fermi-LAT and PAMELA (gray bands) with the extracted size of the $e^\pm$ anomaly (green bands).  Combined statistical and systematic uncertainties are shown for Fermi-LAT and PAMELA $e^-$, while ($\tau = 0.2$) scaled statistical uncertainties are shown for PAMELA $e^+/(e^++e^-)$.  Our background predictions (magenta bands) are also overlaid.}
\label{fig:Backgrounds}
\end{figure}

We use the central values and credibility regions of the parameters determined using the non-$e^\pm$ related data to calculate a background prediction for the $e^\pm$ fluxes. 
Fig. \ref{fig:Backgrounds} shows the the measured $e^\pm$ fluxes and the calculated background.  Statistical and systematic uncertainties combined in quadrature are shown for Fermi-LAT, while ($\tau = 0.2$) scaled statistical uncertainties are shown for PAMELA $e^+/(e^++e^-)$ as gray bands.  Our background prediction is overlaid as magenta bands.  The central value and the 1-$\sigma$ uncertainty of the calculated anomaly is displayed as green dashed lines and bands.  As the first frame shows the Fermi-LAT measurements deviate from the predicted background both below 10 GeV and above 100 GeV.

In our interpretation the deviation is a statistically significant signal of the presence of new physics in the $e^+ + e^-$ flux.  Based on the difference between the central values of the data and the background a similar conclusion can be drawn from PAMELA.  Unfortunately, the sizable uncertainties for the PAMELA measurements prevent us to claim a statistically significant deviation. 
After having determined the background for the $e^\pm$ fluxes, we subtract it from the measured flux to obtain the size of the new physics signal.  Results for the $e^\pm$ anomaly are also shown in Fig. \ref{fig:Backgrounds}.  As expected based on the background predictions a non-vanishing anomaly can be established for the Fermi-LAT $e^+ + e^-$ flux, while no anomaly with statistical significance can be claimed for PAMELA due to the large uncertainties.

\section{The source of the anomaly}

Based on the available evidence we can only speculate about the origin of the discrepancy between the data and predictions of the cosmic electron-positron spectra.  The first obvious assumption is that some aspect of the propagation model used in the present calculation is insufficient for the proper description of the electron-positron fluxes arriving at Earth \cite{Tawfik:2010qf}.  In this case there exists no anomaly in the data.
Assuming that the propagation model satisfactorily describes physics over the Galaxy the next reasonable thing is to suspect local effects modifying the electron-positron distribution \cite{PesceRollins:2009tu}.  Further suspicion falls on the lack of sources included in the calculation \cite{Frandsen:2010mr}.  

Possible new sources of cosmic rays to account for the anomaly have been proposed in two major categories.  The first category is standard astrophysical objects such as supernova remnants, pulsars, various objects in the Galactic centre, etc.  Finally, more exotic explanations call for new astronomical and/or particle physics phenomena, such as dark matter.
In Ref. \cite{Auchettl:2011wi} we compared our extracted signal to recent predictions of anomalous sources.  We considered predictions from supernova remnants, nearby pulsars and dark matter annihilation.  We concluded that presently uncertainties are too large and prevent us from judging the validity of these as explanations of the anomaly.  With more data and more precise calculations the various suggestions of the cosmic $e^- + e^+$ anomaly can be ruled out or confirmed.

\section*{Acknowledgements}

We are indebted to R. Cotta, A. Donea, J. Lazendic-Galloway, Y. Levin and T. Porter for invaluable discussions on various aspects of cosmic ray physics and likelihood analysis.  
CB wishes to thank the organizers and participants of the Bled Workshop for stimulating discussions.
KA is thankful to P. Chan for help with issues of parallel computing, to M. Jasperse for assistance in Mathematica programming, to J. Lazendic-Galloway, T. Porter and to A. Vladimirov for help with GalProp.
This research was funded in part by the Australian Research Council under Project ID DP0877916.  The use of Monash University Sun Grid, a high-performance computing facility, is also gratefully acknowledged.


\begin{thebibliography}{10}
\providecommand{\enquote}[1]{``#1''}
\expandafter\ifx\csname url\endcsname\relax
  \def\url#1{\texttt{#1}}\fi
\expandafter\ifx\csname urlprefix\endcsname\relax\def\urlprefix{URL }\fi

\bibitem{Golden:1992zm}
R.~L. Golden, et~al., \emph{Astrophys. J.} \textbf{436}, 769--775 (1994).

\bibitem{Alcaraz:2000bf}
J.~Alcaraz, et~al., \emph{Phys. Lett.} \textbf{B484}, 10--22 (2000).

\bibitem{Boezio:2001ac}
M.~Boezio, et~al., \emph{Astrophys. J.} \textbf{561}, 787--799 (2001).

\bibitem{Grimani:2002yz}
C.~Grimani, et~al., \emph{Astron. Astrophys.} \textbf{392}, 287--294 (2002).

\bibitem{Barwick:1997ig}
S.~W. Barwick, et~al., \emph{Astrophys. J.} \textbf{482}, L191--L194 (1997).

\bibitem{Beatty:2004cy}
J.~J. Beatty, et~al., \emph{Phys. Rev. Lett.} \textbf{93}, 241102 (2004).

\bibitem{Adriani:2008zr}
O.~Adriani, et~al., \emph{Nature} \textbf{458}, 607--609 (2009).

\bibitem{Timur:2011vv}
T.~Delahaye, F.~Armand, M.~Pohl, and P.~Salati, \emph{arXiv:1102.0744}  (2011).

\bibitem{Aguilar:2002ad}
M.~Aguilar, et~al., \emph{Phys. Rept.} \textbf{366}, 331--405 (2002).

\bibitem{Torii:2008xu}
S.~Torii, et~al., \emph{arXiv:0809.0760}  (2008).

\bibitem{Aharonian:2008aa}
F.~Aharonian, et~al., \emph{Phys. Rev. Lett.} \textbf{101}, 261104 (2008).

\bibitem{Aharonian:2009ah}
F.~Aharonian, et~al., \emph{Astron. Astrophys.} \textbf{508}, 561 (2009).

\bibitem{Ackermann:2010ij}
M.~Ackermann, et~al., \emph{Phys. Rev.} \textbf{D82}, 092004 (2010).

\bibitem{Serpico:2011wg}
P.~D. Serpico, {Astrophysical models for the origin of the positron 'excess'}
  (2011).

\bibitem{Ginzburg:1990sk}
V.~Ginzburg, et~al., \emph{{Astrophysics of cosmic rays}}, {North Holland},
  1990.

\bibitem{Strong:2007nh}
A.~W. Strong, I.~V. Moskalenko, and V.~S. Ptuskin, \emph{Ann. Rev. Nucl. Part.
  Sci.} \textbf{57}, 285--327 (2007).

\bibitem{1994ApJ...431..705S}
E.~S. {Seo}, and V.~S. {Ptuskin}, \emph{Astrophys. J.} \textbf{431}, 705--714
  (1994).

\bibitem{Cotta:2010ej}
R.~C. Cotta, et~al., \emph{JHEP} \textbf{01}, 064 (2011).

\bibitem{DiBernardo:2010is}
G.~Di~Bernardo, et~al., \emph{Astropart.Phys.} \textbf{34}, 528--538 (2011).

\bibitem{Trotta:2010mx}
R.~Trotta, et~al., \emph{The Astrophysical Journal} \textbf{729}, 106 (2010).

\bibitem{Auchettl:2011wi}
K.~Auchettl, and C.~Balazs  (2011), arXiv:1106.4138.

\bibitem{Adriani:2010ib}
O.~Adriani, et~al., \emph{Astropart.Phys.} \textbf{34}, 1--11 (2010).

\bibitem{Adriani:2011xv}
O.~Adriani, et~al., \emph{Phys. Rev. Lett.} \textbf{106}, 201101 (2011).

\bibitem{Adriani:2010rc}
O.~Adriani, et~al., \emph{Phys. Rev. Lett.} \textbf{105}, 121101 (2010).

\bibitem{Moskalenko:2001ya}
I.~V. Moskalenko, et~al., \emph{Astrophys.J.} \textbf{565}, 280--296 (2002).

\bibitem{1988ApJ...328..940K}
K.~E. {Krombel}, and M.~E. {Wiedenbeck}, \emph{Astrophys. J.} \textbf{328},
  940--953 (1988).

\bibitem{1978ApJ...223..676L}
J.~A. {Lezniak}, and W.~R. {Webber}, \emph{Astrophys. J.} \textbf{223},
  676--696 (1978).

\bibitem{1990AA...233...96E}
J.~J. {Engelmann}, et~al., \emph{Astronomy and Astrophysics} \textbf{233},
  96--111 (1990).

\bibitem{PAMELABORONCARBON}
E.~M. et~al., \enquote{{The PAMELA Experiment: Preliminary Results after Two
  Years of Data Taking},} in \emph{21st European Cosmic Ray Symposium (ECRS
  2008)}, 2008, Proceeding of 21st European Cosmic Ray Symposium, pp. 396--401.

\bibitem{Ahn:2008my}
H.~S. Ahn, et~al., \emph{Astropart. Phys.} \textbf{30}, 133--141 (2008).

\bibitem{2000AIPC..528..421D}
A.~J. Davis, et~al., \enquote{{On the low energy decrease in galactic cosmic
  ray secondary/primary ratios},} in \emph{Acceleration and Transport of
  Energetic Particles Observed in the Heliosphere}, edited by {R.~A.~Mewaldt,
  J.~R.~Jokipii, M.~A.~Lee, E.~M{\"o}bius, \& T.~H.~Zurbuchen }, 2000, vol. 528
  of \emph{American Institute of Physics Conference Series}, pp. 421--424.

\bibitem{1999ICRC....3..105H}
M.~{Hareyama}, \enquote{{SUB-Fe/Fe ratio obtained by Sanriku balloon
  experiment},} in \emph{International Cosmic Ray Conference}, 1999, vol.~3 of
  \emph{International Cosmic Ray Conference}, pp. 105--+.

\bibitem{1980ApJ...239L.139W}
M.~E. {Wiedenbeck}, and D.~E. {Greiner}, \emph{The Astrophysical Journal
  Letters} \textbf{239}, L139--L142 (1980).

\bibitem{1981ICRC....9..195G}
M.~{Garcia-Munoz}, et~al., \enquote{{The Energy Dependence of Cosmic-Ray
  Propagation at Low Energy},} in \emph{International Cosmic Ray Conference},
  1981, vol.~9 of \emph{International Cosmic Ray Conference}, pp. 195--+.

\bibitem{2001ICRC....5.1655H}
T.~{Hams}, et~al., \enquote{{$^{10}Be/^{9}Be$ ratio up to 1.0 GeV/nucleon
  measured in the ISOMAX 98 balloon flight},} in \emph{International Cosmic Ray
  Conference}, 2001, vol.~5 of \emph{International Cosmic Ray Conference}, pp.
  1655--+.

\bibitem{davis:421}
A.~J. Davis, et~al., \emph{AIP Conference Proceedings} \textbf{528}, 421--424
  (2000), \urlprefix\url{http://link.aip.org/link/?APC/528/421/1}.

\bibitem{2001AdSpR..27..727Y}
N.~E. Yanasak, et~al., \emph{Advances in Space Research} \textbf{27}, 727--736
  (2001).

\bibitem{2004EPJC...33S.941B2}
J.~{Burger}, \emph{European Physical Journal C} \textbf{33}, 941--943 (2004).

\bibitem{Tawfik:2010qf}
A.~Tawfik, and A.~Saleh, \emph{arXiv:1010.5390}  (2010).

\bibitem{PesceRollins:2009tu}
M.~Pesce-Rollins, and f.~t. F.~L. collaboration, \emph{arXiv:0907.0387}
  (2009).

\bibitem{Frandsen:2010mr}
M.~T. Frandsen, I.~Masina, and F.~Sannino, \emph{arXiv:1011.0013}  (2010).

\end{thebibliography}

\end{document}